\documentclass[]{article}

\usepackage{graphicx}

\usepackage{tikz}
\def\checkmark{\tikz\fill[scale=0.4](0,.35) -- (.25,0) -- (1,.7) -- (.25,.15) -- cycle;} 

\title{\texttt{R} packages for the statistical analysis of doubly truncated data: a review}
\author{Jacobo de U\~{n}a-\'{A}lvarez\\
	{\small Department of Statistics and OR, SiDOR research group \& CINBIO,}\\ {\small University of Vigo, Spain. E-mail: jacobo@uvigo.es}}

\begin{document}

\maketitle

\begin{abstract}
Random double truncation refers a situation in which the variable of interest is observed only when it falls within two random limits. Such phenomenon occurs in many applications of Survival Analysis and Epidemiology, among many other fields. There exist several \texttt{R} packages to analyze doubly truncated data which implement, for instance, estimators for the cumulative distribution, the cumulative incidences of competing risks, or the regression coefficients in the Cox model. In this paper the main features of these packages are reviewed. The relative merits of the libraries are illustrated through the analysis of simulated and real data. This includes the study of the statistical accuracy of the implemented techniques as well as of their computational speed. Practical recommendations are given.\\
%\medskip

%The nonparametric maximum-likelihood estimator of a doubly truncated distribution has no explicit form and must be computed iteratively. In this paper we introduce the package \texttt{DTDA.cif} which implements numerical algorithms to estimate the distribution function of a doubly truncated variable as well as the cumulative incidence functions attached to a categorical covariate. The latter are of much interest in the analysis of competing events or, more generally, when the focus is on the subdistribution functions induced by a particular covariate. The package offers two different estimation approaches depending on whether or not the truncation limits depend on the covariate. Standard errors based on the simple bootstrap and confidence intervals are available too. Practical and computational advantages with respect to existing software for random double truncation are pointed out.
\end{abstract}

\section{Introduction}

Doubly truncated data appear in fields like Astronomy, Survival Analysis, Epidemiology, Engineering or Economics, among others. A random variate $X$ is doubly truncated by a random couple $(U,V)$ when the observations are limited to the triplets $(X,U,V)$ which satisfy $U \leq X \leq V$. In such a setting the variables $U$ and $V$ are called the left and right truncation limits for $X$. Efron and Petrosian (1999) considered the estimation of the cumulative distribution function (cdf) of quasar luminosities (the $X$), which go undetected whenever they fall outside or region $[U,V]$; thus, the $X$ is doubly truncated by $(U,V)$. Double truncation occurs with interval sampling too, which in particular relates the analysis of epidemiological registries. For instance, Moreira and de U\~{n}a-\'{A}lvarez (2010) estimated the cdf of the age at onset for childhood cancer (the $X$) from the cases diagnosed between two specific dates $d_0$ and $d_1$. Then, the sample consisted of the children with $U\leq X\leq V$, where $V$ is the age by $d_1$ and $U=V-d_0$. See also Zhu and Wang (2014), Ye and Tang (2016), D\"{o}rre (2017) or Mandel et al. (2018) for more applications with interval sampling.\\

The double truncation phenomenon can be seen as a generalization of one-sided (left or right) truncation, which has been extensively investigated in the literature, and which requires proper corrections in order to eliminate the sampling bias. However, unlike for one-sided truncation, the nonparametric maximum likelihood estimator (NPMLE) for doubly truncated data has no explicit form and must be computed iteratively. This has motivated the development of software routines which, indeed, have proliferated in the recent years. Specifically for \texttt{R} software (R Core Team, 2017), at the present date at least five different libraries to estimate the cdf of a doubly truncated variable exist; they include some other functionalities too. The methods implemented by these libraries are heterogeneous in both precision and computational speed. A review of these packages is important in order to (i) clearly describe which methods are available in \texttt{R} to analyze doubly truncated data; (ii) how is the relative performance of the several existing libraries in terms of statistical accuracy and execution time; and (iii) give practical recommendations to the scientific community. The aim of this piece of work is to achieve these goals.\\ 

The rest of the paper is organized as follows. In Section 2 several algorithms to compute the NPMLE of the cdf of interest and its standard error are reviewed. Estimation approaches for Cox regression are discussed too. Section 3 briefly describes the main features of the existing \texttt{R} packages to analyze doubly truncated data, while Section 4 investigates through simulations their relative merits, including computational speed and statistical accuracy. In Section 5 several illustrative real data applications are given; these concern the age at diagnosis of childhood cancer (together with cumulative incidences for the several cancer types) as well as the age at onset of Parkinson's disease and its relationship with single nucleotide polymorphisms. A final discussion and some practical recommendations are reported in Section 6.

\section{Estimation approaches and algorithms}

Efron and Petrosian (1999) introduced the NPMLE of the cdf from doubly truncated data, as the maximizer of the conditional likelihood of the $X_i$'s given the $(U_i,V_i)$'s. Here $(X_i,U_i,V_i)$, $1\leq i\leq n$, denotes a random sample with the conditional distribution of $(X,U,V)$ given $U\leq X\leq V$; the variables $X$ and $(U,V)$ are assumed to be independent. Two iterative procedures to compute the NPMLE were proposed in the aforementioned paper. The first algorithm is based on a self-consistency equation; the second one updates the Lynden-Bell estimator for left-truncated data by improving the hazard rate step by step. Efron and Petrosian (1999) recommended the second algorithm when the right truncation is light due to its relative convergence speed.\\

Shen (2010) showed that Efron and Petrosian (1999)'s conditional NPMLE indeed maximizes the full likelihood of the $(X_i,U_i,V_i)$'s. Besides, he proposed a method to jointly estimate the cdf of $X$, $F$ say, and that of $(U,V)$, $K$ say. Specifically, the method iterates by using the couple of equations

\begin{itemize}
	\item [(S1)] $F_n(x)=\sum_{i=1}^n w_i(G_n)I(X_i \leq x) / \sum_{i=1}^n w_i(G_n)$,
	\item [(S2)] $K_n(u,v)=\sum_{i=1}^n w_i(F_n) I(U_i\leq u,V_i\leq v) / \sum_{i=1}^n w_i(F_n)$
	
\end{itemize}

\noindent where $w_i(G_n)=G_n(X_i)^{-1}$ with $G_n(x)=K_n(X_i,\infty)-K_n(X_i,X_i-)$, and where $w_i(F_n)=(F_n(V_i)-F_n(U_i-))^{-1}$. One advantage of (S1)--(S2) is that, unlike the algorithms in Efron and Petrosian (1999), it reconstructs the truncation distribution from the data and, from this, the sampling probabilities for the $X$-values $G(X_i)=P(U\leq X_i\leq V)=K(X_i,\infty)-K(X_i,X_i-)$ can be evaluated. These sampling probabilities play a critical role in the development of regression methods for doubly truncated data, as it will become clear soon. The asymptotic properties of $F_n$ were investigated by Shen (2010); the results and proofs in that paper have however some flaws that were corrected only recently, see de U\~{n}a-\'{A}lvarez and Van Keilegom (2019).\\

Mandel et al. (2018) and, independently, Rennert and Xie (2018) introduced two different estimation approaches for the Cox model $h(x|z)=h_0(x)\exp(\beta z)$ under double truncation. The estimator of Mandel et al. (2018) solves the score equation

$$U(\beta)\equiv \sum_{i=1}^n [Z_i-\frac{\sum_{j=1}^n Z_j \exp(\beta Z_j) G_n(X_j)^{-1}I(X_j\geq X_i)}{\sum_{j=1}^n \exp(\beta Z_j) G_n(X_j)^{-1}I(X_j\geq X_i)}]=0,$$

\noindent where $Z_i$ is the covariate vector attached to $(X_i,U_i,V_i)$. On the other hand, Rennert and Xie (2018) investigated an alternative approach in which the $i$-th term in $U(\beta)$ is weighted by $G_n(X_i)^{-1}$. This extra weighting may introduce more variance in estimation, a fact which is explored in Section 4. Both estimation approaches rely on the independence between the truncation variables and the target $(X,Z)$. The weights $G_n(X_i)^{-1}$ were used in de U\~{n}a-\'{A}lvarez (2020) to introduce cumulative incidence functions for competing risks too. The method is consistent when $(U,V)$ is independent of the competing event. When the distribution of $(U,V)$ varies along the competing risks, the sampling probabilities $G(X_i)$ must be separately estimated from the several groups of events. This however may result in very noisy estimates, particularly when few events are recorded. The theoretical validity of the estimation approaches for both Cox regression and competing risks relies on the asymptotic theory developed in de U\~{n}a-\'{A}lvarez and Van Keilegom (2019).\\

One important issue is that of the estimation of standard errors. For $F_n(x)$, both the simple and the obvious bootstrap have been shown to perform well (Moreira and de U\~{n}a-\'{A}lvare, 2010). The simple bootstrap was compared to the Jackknife approach and to an explicit-form variance estimator in Emura et al. (2015); see Section 4.1 below for an independent simulation study. No theoretical, rigurous proof of consistency has been provided so far for any of the existing estimators of the standard error . Also interestingly, Hu and Emura (2015) and Emura et al. (2017) proposed estimation and inference techniques in the setting in which $F$ is assumed to belong to a parametry family of cdf's. Finally, we mention that a non-iterative approximation to the NPMLE $F_n$ was introduced by de U\~{n}a-\'{A}lvarez (2018).\\

%To mention: interval sampling, quasi-independence assumption, identifiability issues (support conditions), existence and uniqueness of the NPMLE.\\

%To be included: the self-consistency algorithm of fron and Petrosian (1999); the Lynden-Bell-based algorithm (same reference); and the numerical solution for the score equations in Shen (2010). Also, the non-iterative estimator for interval sampling.\\

%To be discussed: bootstrap resampling algorithms (simple, obvious) and bootstrap confidence intervals. Emura's alternative estimator for the standard error.\\

%Any word on Cox regression, on testing for quasi-independence, or on parametric models?

\section{Existing \texttt{R} packages}

In this Section the main features of five existing \texttt{R} packages to analyze doubly truncated data are briefly reviewed. Some of these packages include automatic plots too but such capabilities are not discussed here. The features of the packages are summaryzed in Table \ref{tab:packages}.

\subsection{Package \texttt{DTDA}}

The package \texttt{DTDA} (Moreira et al., 2020) was launched in September 2009, when the statistical community was still quite unaware of the relevance of double truncation. \texttt{DTDA} was the first \texttt{R} library implementing numerical algorithms to compute the NPMLE for doubly truncated data, and it became popular very soon. Last version 2.1-2 (February 2020) introduced some corrections. The main functions in \texttt{DTDA} are \texttt{efron.petrosian}, \texttt{lynden} and \texttt{shen}, implementing respectively the self-consistency algorithm of Efron and Petrosian (1999), the Lynden-Bell-based algorithm, and a numerical solution for the score equations in Shen (2010). By default 95\% confidence limits for the marginal cdfs of the variable of interest and (in the case of \texttt{shen}) of the truncating variables are calculated; for this, the simple bootstrap with 500 replicates (default) and the percentile method are used. The obvious bootstrap, which requires the preliminary estimation of the truncation distribution, is available for \texttt{shen} too. The confidence level is an argument of the main functions and therefore its default value of 95\% can be changed. See Moreira et al. (2010) for more details.

\subsection{Package \texttt{SurvTrunc}}

Lior Rennert introduced in July 2018 his package \texttt{SurvTrunc} (Rennert, 2018), including the function \texttt{cdfDT} which implements the solution to Shen (2010)'s score equations. This function allows for the computation standard errors and of 95\% confidence limits for the cdf of interest based on the simple bootstrap and the normal approximation method; the confidence level is fixed. Confidence limits for the truncation distribution are not implemented. Package \texttt{SurvTrunc} includes the function \texttt{CoxDT} which allows for Cox regression under double truncation too, based on Rennert and Xie (2018). Finally, the function \texttt{indeptestDT} performs a test for the quasi-independence between the variable of interest and the truncating variables, based on the conditional Kendall's Tau of Martin and Betensky (2005).

\subsection{Package \texttt{double.truncation}}

Simultaneously with, and independently of, \texttt{SurvTrunc}, in July 2018 Takeshi Emura launched the package \texttt{double.truncation} (Emura et al., 2019), implementing several parametric models for likelihood-based inference (Hu and Emura, 2015; Emura et al., 2017). The package was last updated by January 2019 (version 1.4). For nonparametric inference, the function \texttt{NPMLE} was included. This function implements the self-consistency algorithm of Efron and Petrosian (1999) as well as an explicit-form estimator for the standard error of the NPMLE introduced in Emura et al. (2015); this standard error does not require any bootstrapping.

\subsection{Package \texttt{DTDA.cif}}

In October 2019 a new package to analyse doubly truncated data appeared: \texttt{DTDA.cif} (de U\~{n}a-\'{A}lvarez and Soage, 2020). Later updated in February 2020, \texttt{DTDA.cif} allows for nonparametric estimation of cumulative incidence functions for competing risks. The main function is \texttt{DTDAcif}, which implements two different estimators together with standard errors based on the simple bootstrap. One of the available estimators (\texttt{method="indep"}) is recommended when the truncation variables are independent of the competing events, while the other one (\texttt{method="dep"}) corrects for possible dependences between the truncation mechanism and the event indicator; see de U\~{n}a-\'{A}lvarez (2020) for details. When applied to a single type of event, function \texttt{DTDAcif} returns a numerical solution to the score equations of Shen (2010) in the spirit of \texttt{shen\{DTDA\}}. Even in the standard setting without competing risks, \texttt{DTDA.cif} is of interest, since it offers computational advantages compared to \texttt{DTDA}. See Section 4.1 for more on this.

\subsection{Package \texttt{DTDA.ni}}

A non-iterative estimator of the cdf under double truncation was introduced by de U\~{n}a-\'{A}lvarez (2018). This is a nonparametric estimator which can be computed in $n$ steps (no numerical approximation is needed) and which, in the simulations studies in the referred article, performs almost as well as the Efron-Petrosian NPMLE. The non-iterative estimator was implemented in the main function \texttt{DTDAni} of the package \texttt{DTDA.ni} (de U\~{n}a-\'{A}lvarez and Soage, 2018), launched in April 2018. The implementation is restricted to interval sampling. No method to approximate the standard error is available within the package.

\begin{table}[!ht]\centering
	\caption{\label{tab:packages} Summary of features for the several \texttt{R} functions. Focus is on estimation and inference for $F$; other features of the functions/libraries are reported too. Labels indicate availability of point estimates (Est), standard errors (Std.err) and confidence intervals (CI)}
	\medskip
	\begin{tabular}{lcccl}
		% First line:
		%	\toprule[0.09 em]
		% The body of the table:
		\hline
		 function/libary & Est & Std.err & CI & other features\\
		\hline
		%	\midrule
		 \texttt{efron.petrosian\{DTDA\}}  & \checkmark & -- & \checkmark & simple boot (\texttt{B=500}) \\
		 \texttt{lynden\{DTDA\}}  & \checkmark & -- & \checkmark & simple boot (\texttt{B=500})\\ %
		 \texttt{shen\{DTDA\}}  & \checkmark & -- & \checkmark & simple/obvious boot (\texttt{B=500})\\
				  &  &  &  & estimates for $K$ and $G$\\
		 \texttt{cdfDT\{SurvTrunc\}}  & \checkmark & \checkmark & \checkmark & simple boot (\texttt{B=200})\\
		  &  &  &  & estimate for $K$\\
		 &  &  &  & Cox model, quasiindep test\\
		 \texttt{NPMLE\{double.truncation\}}  & \checkmark & \checkmark & -- & explicit-form std err\\ 
		 		  &  &  &  & parametric models\\%
		 \texttt{DTDAcif\{DTDA.cif\}}  & \checkmark & \checkmark & -- &  simple boot (\texttt{B=300})\\
		 	 &  &  &  & estimate for $G$\\
		 	 	 &  &  &  & cumulative incidences\\
		 \texttt{DTDAni\{DTDA.ni\}}  & \checkmark & -- & -- &  non-iterative estimator\\
		% last line:
		%	\bottomrule[0.09 em]
		\hline
	\end{tabular}
\end{table}

\section{Relative performance of the packages}

In this Section the relative performance of the packages listed in Section 3 is investigated through simulated data. More specifically, the focus will be on the execution time when computing the NPMLE (or the non-iterative estimator) and the pertaining standard errors. Also, attention will be paid to the relative accuracy of the two existing estimation approaches for the Cox regression model. 

\subsection{Computational time for the NPMLE}

The execution time for the several existing implementations of the NPMLE is variable, and the differences can be very remarkable. Some of the \texttt{R} functions only compute the estimator for $F$, the cdf of interest; these are \texttt{efron.petrosian\{DTDA\}}, \texttt{lynden\{DTDA\}} and \texttt{NPMLE\{double.truncation\}}. On the other hand, \texttt{shen\{DTDA\}}, \texttt{cdfDT\{SurvTrunc\}} and \texttt{DTDAcif\{DTDA.cif\}} calculate in a simultaneous way the estimator of $F$ and that of the truncation distribution, leading to larger waiting times. Also, when computing standard errors or confidence limits an extra computational effort is needed. This is particularly important for the bootstrap-based methods. Finally, the sample size $n$ may have a role in the relative computational speed too. All these aspects are illustrated now through simulations.\\

\begin{figure}[htp]
	\centering	

	%\caption{equation...}	
	\begin{tabular}{cc}		
		\includegraphics[width=60mm]{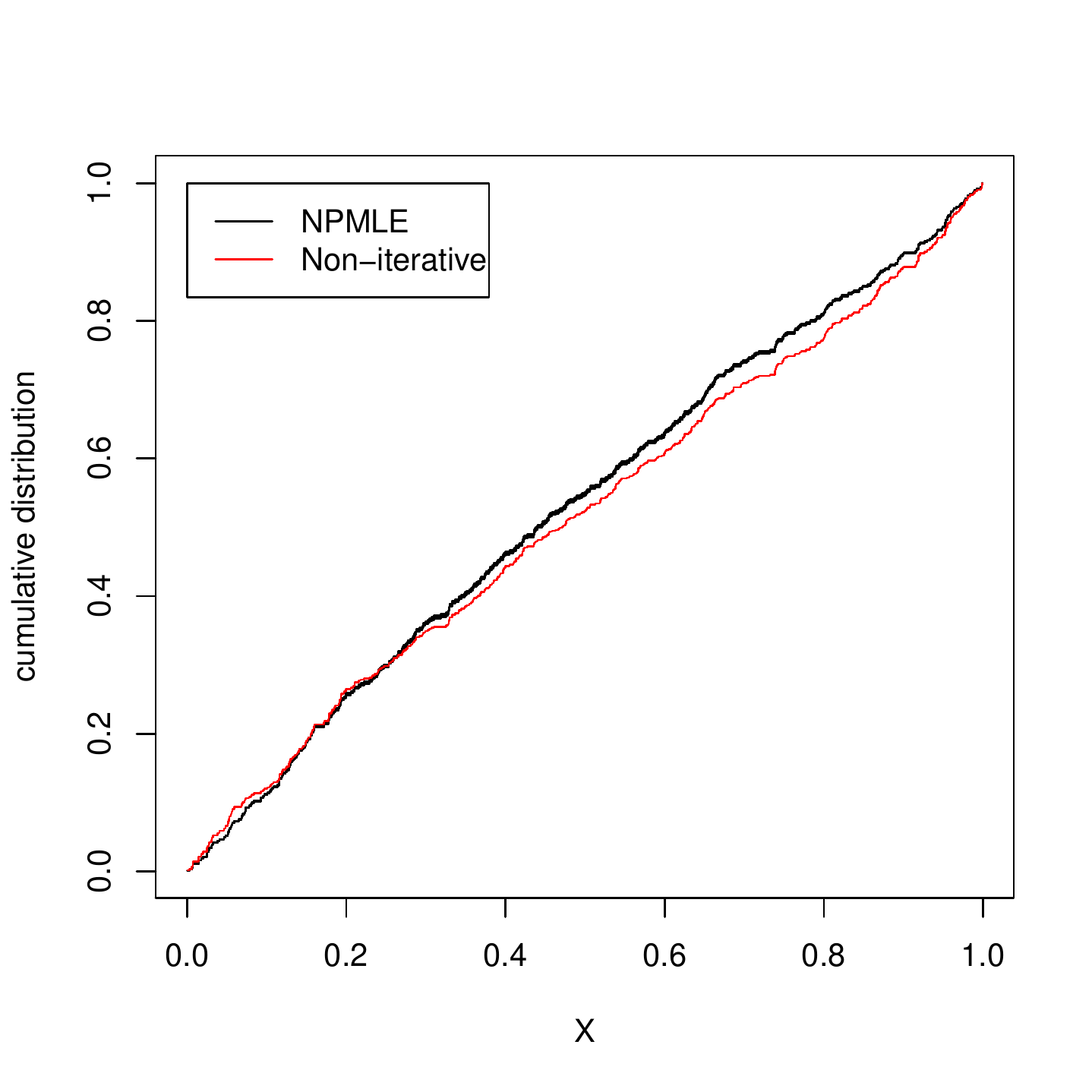}&	
		\includegraphics[width=60mm]{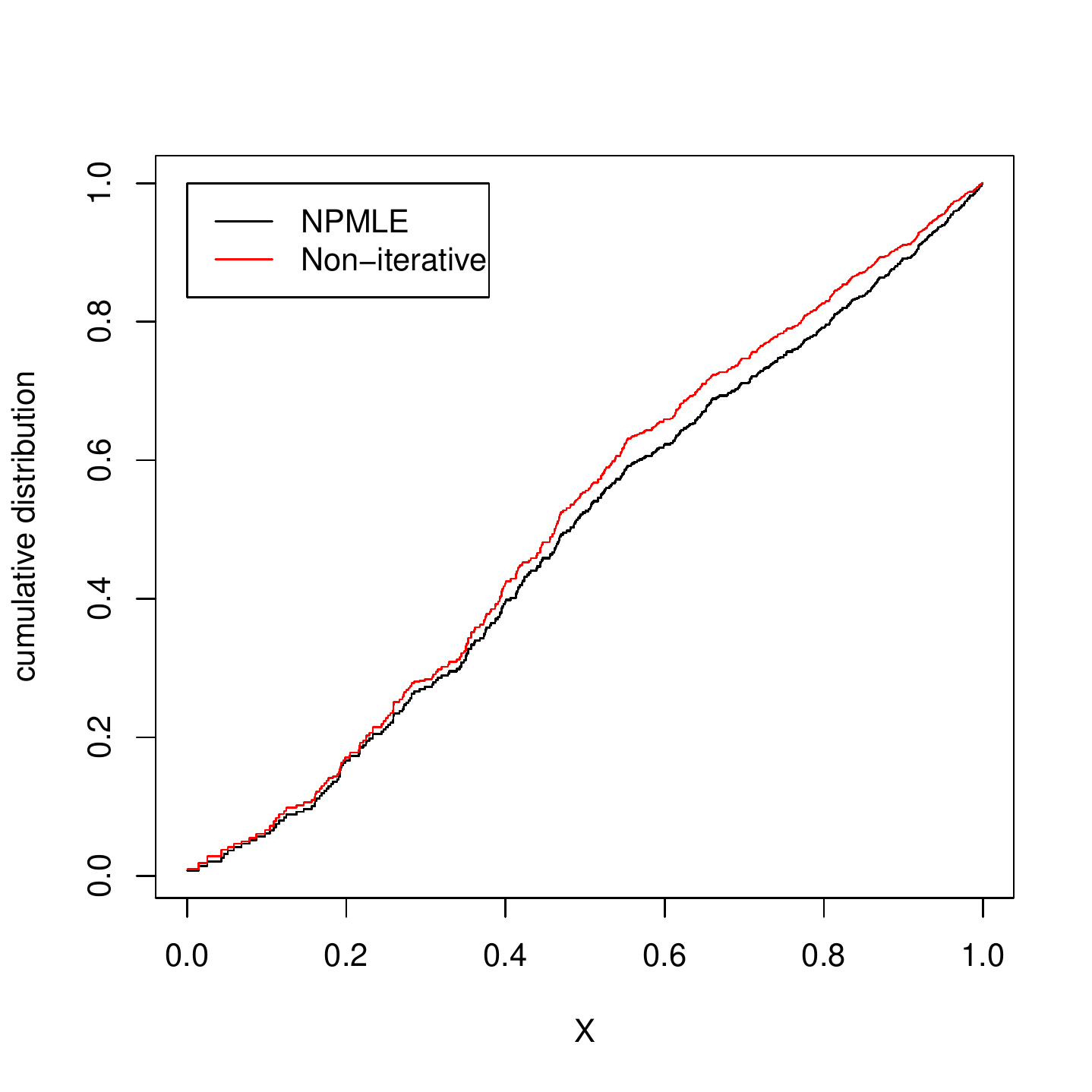}\\		
	\end{tabular}
	\caption{NPMLE (black line) and non-interative estimator (red line) computed from a single sample of size $n=500$ from the simulated model. Left: $\rho=1$; right: $\rho=0.5$. The true cdf is uniform on the interval $(0,1)$.}	
		\label{fig: exec_noboot}
\end{figure}

A variable $X$ uniformly distributed on the $(0,1)$ interval was simulated, $X \sim U(0,1)$. Random double truncation was introduced by simulating model (LT)-(RT) in Section 4.2, with $\rho=1$ or $\rho=0.5$ and with $\tau=0.25$. The running times for the several existing implementations of the NPMLE, as well as for the non-iterative estimator, were computed. To this end, a single sample with size $n$ ($n=250$, $500$ or $1,000$) was drawn from the model. These execution times are provided in Table \ref{tab:exec_noboot}. For illustration purposes, the resulting estimators for $n=500$ are displayed in Figure \ref{fig: exec_noboot}; the six \texttt{R} functions which implement the NPMLE report roughly the same result, while the non-iterative estimator returned by \texttt{DTDAni} is only slightly different to the NPMLE.\\

\begin{figure}[htp]
	\centering	

	%\caption{equation...}	
	\begin{tabular}{cc}		
		\includegraphics[width=60mm]{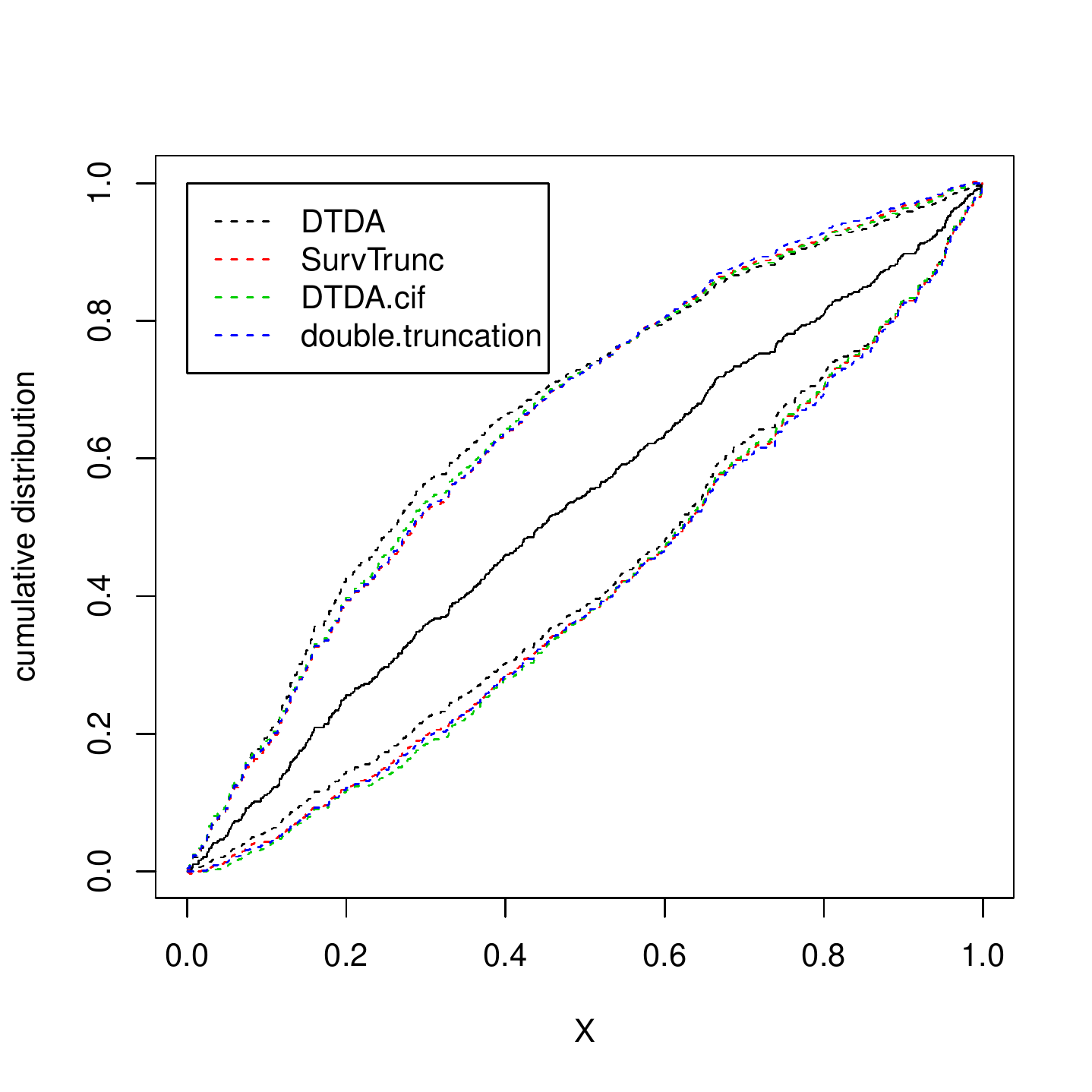}&	
		\includegraphics[width=60mm]{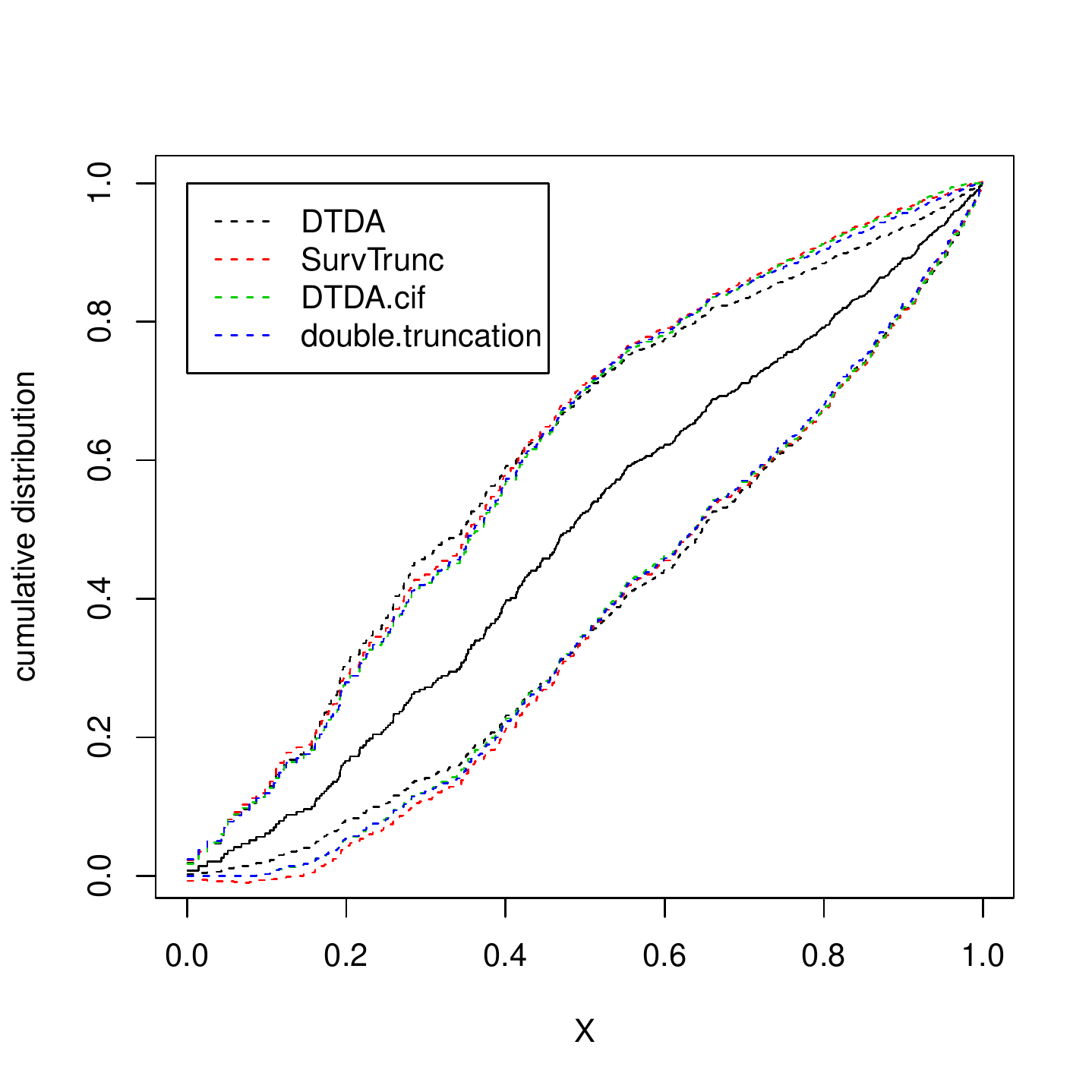}\\		
	\end{tabular}
	\caption{NPMLE (solid line) and pointwise 95\% confidence limits (dashed lines) computed from a single sample of size $n=500$ from the simulated model. Left: $\rho=1$; right: $\rho=0.5$. The true cdf is uniform on the interval $(0,1)$.}	
		\label{fig: exec_boot}
\end{figure}

From Table \ref{tab:exec_noboot} it is seen that the smallest running times for the computation of the NPMLE correspond to \texttt{efron.petrosian}, the second best being \texttt{DTDAcif}. Function \texttt{NPMLE} is quite competitive, but its results worsen for $n=1,000$, being about 5 times slower than \texttt{efron.petrosian}. This could be due to the computation of standard errors, which is performed by default and cannot be deactivated. On the other hand, the running time for \texttt{cdfDT} is about 3 times that of \texttt{DTDAcif}. Although the former returns the NPMLE for the truncating distribution, which is not the case for the latter, the function \texttt{DTDAcif} saves the sampling probabilities $G_n(X_i)$, $1 \leq i \leq n$, which is the essential piece of information on the truncating variables for most applications. Finally, \texttt{lynden} performs badly, particularly for the largest sample size, while the elapsed times for \texttt{shen} are just inadmissible. Summarizing, the fastest option to compute the NPMLE is \texttt{efron.petrosian}, but \texttt{DTDAcif} is recommended when the sampling probabilties are to be computed too. Regarding standard errors, an additional study was performed, including the calculation of bootstrap standard errors (or bootstrap confidence limits) for \texttt{efron.petrosian}, \texttt{cdfDT} and \texttt{DTDAcif}. We took 99 bootstrap resamples in all the cases; obviously, in general the running times in Table \ref{tab:exec_noboot} will be multiplied by the resampling effort. The main conclusions of this additional study are:

\begin{itemize}
	\item [(a)] \texttt{NPMLE} is by far the fastest for the computation of standard errors; for example, with $n=500$, \texttt{NPMLE} took 2 seconds compared to the about 60 seconds of \texttt{efron.petrosian}, which was the quickest option when bootstrapping
	\item [(b)] there exists a big difference in computational speed among the three functions implementing the bootstrap; for example, with $n=250$ the execution times were about 11, 65 and 25 seconds for \texttt{efron.petrosian}, \texttt{cdfDT} and \texttt{DTDAcif}, respectively (in agreement with Table \ref{tab:exec_noboot})
	\item [(c)] the 95\% confidence limits based on the (simple) bootstrap and on the explicit-form standard error implemented in \texttt{NPMLE\{double.truncation\}} were quite similar for the simulated trials; see Figure \ref{fig: exec_boot} for an illustration in the case $n=500$
\end{itemize}

\noindent One last comment on Table \ref{tab:exec_noboot} is that the execution time for the non-iterative estimator was very fast compared to the NPMLE. This was somehow expected since the former does not require any iterations.\\

\begin{table}[t!]\centering
	%	\medskip
	\begin{tabular}{ccccccc}
		% First line:
		%	\toprule[0.09 em]
		% The body of the table:
		\hline
		& & $\rho=1$ &  & & $\rho=0.5$  & \\
		\hline
		& $n=250$ & $n=500$ & $n=1,000$   & $n=250$ & $n=500$ & $n=1,000$  \\
		\hline
		%	\midrule
		\texttt{efron.petrosian}  & 0.11 & 0.40 & 2.14 & 0.12 & 0.45 & 2.37 \\
		\texttt{lynden}  &  6.10 & 12.10 & 52.39 & 2.70 & 10.70 & 65.04 \\ %
		\texttt{shen}  & 8.55  & 90.92 & 565.48 & 8.53 & 112.74 & 587.34 \\
		\texttt{cdfDT}  & 0.75  & 2.96 & 6.89 & 0.64 & 3.08 & 6.51 \\
		\texttt{NPMLE}  & 0.32  & 1.95 & 10.18 & 0.30 & 2.09 & 10.09 \\ %
		\texttt{DTDAcif}  & 0.15  & 0.81 & 2.63 & 0.12 & 0.86 & 2.69 \\
		\texttt{DTDAni}  &  0.02 & 0.14 & 0.61 & 0.03 & 0.20 & 0.88 \\
		% last line:
		%	\bottomrule[0.09 em]
		\hline
	\end{tabular}
	\caption{\label{tab:exec_noboot} Execution time (seconds) for the several existing implementations of the NPMLE and for the non-iterative estimator. One single sample of simulated data with size $n$.}
\end{table}

An interesting issue is that of the statistical accuracy of the standard error implemented in \texttt{NPMLE\{double.truncation\}} relative to the bootstrap standard error reported for instance by \texttt{DTDAcif}. Note that the latter requires a quite larger execution time. In Table \ref{tab:stderr_M250} the bias, standard deviation (SD) and mean squared error (MSE) of the estimated standard error of $F_n(x)$ along $250$ Monte Carlo trials are reported. The distribution of $X$ was $U(0,1)$ and the truncation model followed (LT)-(RT) in Section 4.2. The chosen values for $x$ were the three quartiles of $X$. The number of bootstrap resamples were $99$ and the sample sizes were $50$, $100$ and $250$. The true standard error of $F_n(x)$ (the target) was approximated by the Monte Carlo standard deviation from an independent experiment with $1,000$ trials.\\

From Table \ref{tab:stderr_M250} it is seen that the simple bootstrap implemented in \texttt{DTDAcif} behaved better than the explicit-form standard error for moderate sample sizes ($n=50,100$). This was mainly due to the relative smaller standard deviation of the bootstrap. Indeed, the explicit-form estimator reported several outliers, leading to the aforementioned relative large dispersion and to a visible positive bias ($n=50$) too. Relative results were the opposite for $n=250$, although the bootstrap was competitive in this case too. Interestingly, in the case $n=250$ the issue of \texttt{NPMLE} of having outliers just disappeared. The conclusions on the relative mean square error are in agreement with the independent simulation study in Emura et al. (2015).\\

The method implemented in \texttt{NPMLE} reported \texttt{NaN} results in a number of trials, with a warning related to \texttt{sqrt(V\_F)}. For instance, in the case $n=100$ the standard error of $F_n(x)$ was not available because of this reason for $15$ ($x=x_{.25}$), $10$ ($x=x_{.5}$) and $8$ ($x=x_{.75}$) trials out of the 250. The situation was worse for $n=50$ when, besides, an error in \texttt{solve.default(Info)} was declared for 8\% of the trials; these trials were eliminated in the summaries of Table \ref{tab:stderr_M250}. Overall, with $n=50$ the percentage cases for which \texttt{NPMLE} did not report the standard error ranged from $17$\% ($x=x_{0.75}$) to $28$\% ($x=x_{0.25}$). \\

%This is why results for \texttt{NPMLE} are not reported in Table \ref{tab:stderr} for the case $n=50$.\\

%\begin{table}[!ht]\centering
%	\caption{\label{tab:stderr} Bias, standard deviation (SD) and mean squared error (MSE) along $100$ trials of estimated standard errors for $F_n(x)$: simple bootstrap based on $99$ resamples as computed by \texttt{DTDAcif}, and explicit-form estimator implemented by \texttt{NPMLE}. The chosen values for $x$ are the three quartiles of $X$.}
%	\medskip
%	\begin{tabular}{ccccccccccc}
		% First line:
		%	\toprule[0.09 em]
		% The body of the table:
%		\hline
%		& $x$ & & $n=50$ & & & $n=100$ & & & $n=250$ & \\
%		\hline
%		& & Bias & SD & MSE   & Bias & SD & MSE  & Bias & SD & MSE \\
%		\hline
%		%	\midrule
%		\texttt{DTDAcif} & $x_{.25}$ & -0.0435 & 0.0979 & 0.0115 & -0.0030 & 0.0792 & 0.0063 &  0.0004 &  0.0395  & 0.0016 \\
%	                     & $x_{.5}$ & -0.0178 & 0.0763 & 0.0061 & -0.0085 & 0.0562 & 0.0032 & -0.0015 & 0.0215 & 0.0005 \\
%		                 & $x_{.75}$ &  -0.0029 & 0.0758 &  0.0058 & -0.0080 & 0.0484 &  0.0024 & 0.0024 & 0.0194 & 0.0004 \\
%		\texttt{NPMLE}  & $x_{.25}$ & 0.0659 &  0.6415 & 0.4158 & -0.0624 & 0.0877 & 0.0116 &  -0.0078 & 0.0348 & 0.0013\\ 
%		                & $x_{.5}$ & 0.0132 & 0.4257 & 0.1814 & -0.0364 & 0.0587 & 0.0048 &  -0.0018 &  0.0175 &  0.0003 \\ 
%		                & $x_{.75}$ & 0.0025 &  0.2266 & 0.0513 &  -0.0219 & 0.0475 & 0.0027 & 0.0024 & 0.0156 & 0.0002 \\ 
		% last line:
		%	\bottomrule[0.09 em]
%		\hline
%	\end{tabular}
%\end{table}

\begin{table}[t!]\centering
	%\medskip
	\begin{tabular}{cccccccccc}
		% First line:
		%	\toprule[0.09 em]
		% The body of the table:
		\hline
		$x$ & & $n=50$ & & & $n=100$ & & & $n=250$ & \\
		\hline
		& Bias & SD & MSE   & Bias & SD & MSE  & Bias & SD & MSE \\
		\hline
		%	\midrule
		& & & & &	\texttt{DTDAcif} & & & & \\
		$x_{.25}$ & -0.0416 & 0.0996 & 0.0117 & -0.0001 & 0.0814 & 0.0066 & 0.0038 & 0.0392 & 0.0016\\
		$x_{.5}$ & -0.0189 & 0.0719 & 0.0055 & -0.0035 & 0.0530 & 0.0028 & -0.0016 & 0.0211 & 0.0004\\
		$x_{.75}$ & -0.0098 & 0.0761 & 0.0059 &  -0.0047 & 0.0475 & 0.0023 & 0.0004 & 0.0190 & 0.0004\\
		&  & & & & \texttt{NPMLE} & & & & \\ 
		$x_{.25}$ & 0.2658 & 1.5794 & 2.5651 & -0.0240 & 0.2762 & 0.0768 & -0.0051 & 0.0354 & 0.0013\\ 
		$x_{.5}$ & 0.1819 & 1.3800 & 1.9376 & -0.0081 & 0.1798 & 0.0324 & -0.0016 & 0.0186 & 0.0004\\ 
		$x_{.75}$ & 0.1239 & 1.2324 & 1.5340 &  -0.0065 & 0.1040 & 0.0109 & 0.0002 & 0.0185 & 0.0003\\ 
		% last line:
		%	\bottomrule[0.09 em]
		\hline
	\end{tabular}
	\caption{\label{tab:stderr_M250} Bias, standard deviation (SD) and mean squared error (MSE) along $250$ trials of estimated standard errors for $F_n(x)$: simple bootstrap based on $99$ resamples as computed by \texttt{DTDAcif}, and explicit-form estimator implemented by \texttt{NPMLE}. The chosen values for $x$ are the three quartiles of $X$.}
\end{table}

%Comments on: statistical accuracy of Emura's explicit-form standard error; bootstrapping the non-iterative estimator\\

%Comments on: computational times; accuracy of explicit-form relative to bootstrap-based standard errors; accuracy of Rennert and Xie's Cox regression relative to Mandel et al.'s; non-iterative estimator vs. Efron-Petrosian NPMLE\\

%Simulations: are the methods doing it well?

\subsection{Cox regression}

Cox regression with a doubly truncated response can be performed in \texttt{R} by including the estimated, transformed sampling probabilities as an \texttt{offset} in \texttt{coxph}. Specifically, if the $(X_i,Z_i,U_i,V_i)$ are respectively saved in \texttt{x}, \texttt{z}, \texttt{u} and \texttt{v}, the code lines

\begin{verbatim}
> W <- DTDAcif(x, u, v)$biasf
> coef(coxph(Surv(x) ~ z + offset(-log(W))))
\end{verbatim}

\noindent return the estimator of the regression coefficients proposed in Mandel et al. (2018). Note that the object \texttt{W} contains the values $G_n(X_i)$, $1 \leq i\leq n$. As mentioned above, Mandel et al. (2018)'s method is different to the estimator in Rennert and Xie (2018); the latter is implemented in the function \texttt{coxDT} of the package \texttt{SurvTrunc}, and is returned by

\begin{verbatim}
> coxDT(Surv(x) ~ z, u, v, data = sim.d, B.SE.np = 2)$results.beta[1]
\end{verbatim}

In order to compare both estimation approaches, data with samples sizes $n=250$, $500$ and $1,000$ from a Cox regression model were simulated. The simulated model was

\begin{equation}
h(x|z)=h_0(x) \exp(\beta z)
\end{equation}

\noindent with baseline hazard $h_0(x)=(1/\sigma)x^{1/\sigma-1}$ and regression coefficient $\beta=1/\sigma$, so the conditional distribution of $X$ given $Z=z$ is Weibull with shape parameter $1/\sigma$ and scale parameter $\exp(-z)$. This corresponds to a loglinear regression model $\log(X)=-Z+\sigma \epsilon$, where the error term $\epsilon$ follows a extreme value distribution independent of $Z$. We took $\sigma=0.1$ so the regression coefficient was $\beta=10$. The chosen model for $Z$ was exponential with rate 1. In this scenario the essential support of $X$ falls within the interval $(0,1.11)$. In order to introduce double truncation, the following model was considered:

\begin{itemize}
	\item [(LT)] $U=(1+\tau)\xi^\rho-\tau$, $\xi \sim U(0,1)$, $\tau=0.25$, $\rho=0.5$
	\item [(RT)] $V=U+\tau$
\end{itemize}

\begin{figure}[h]
	\centering
	\includegraphics[scale=0.7]{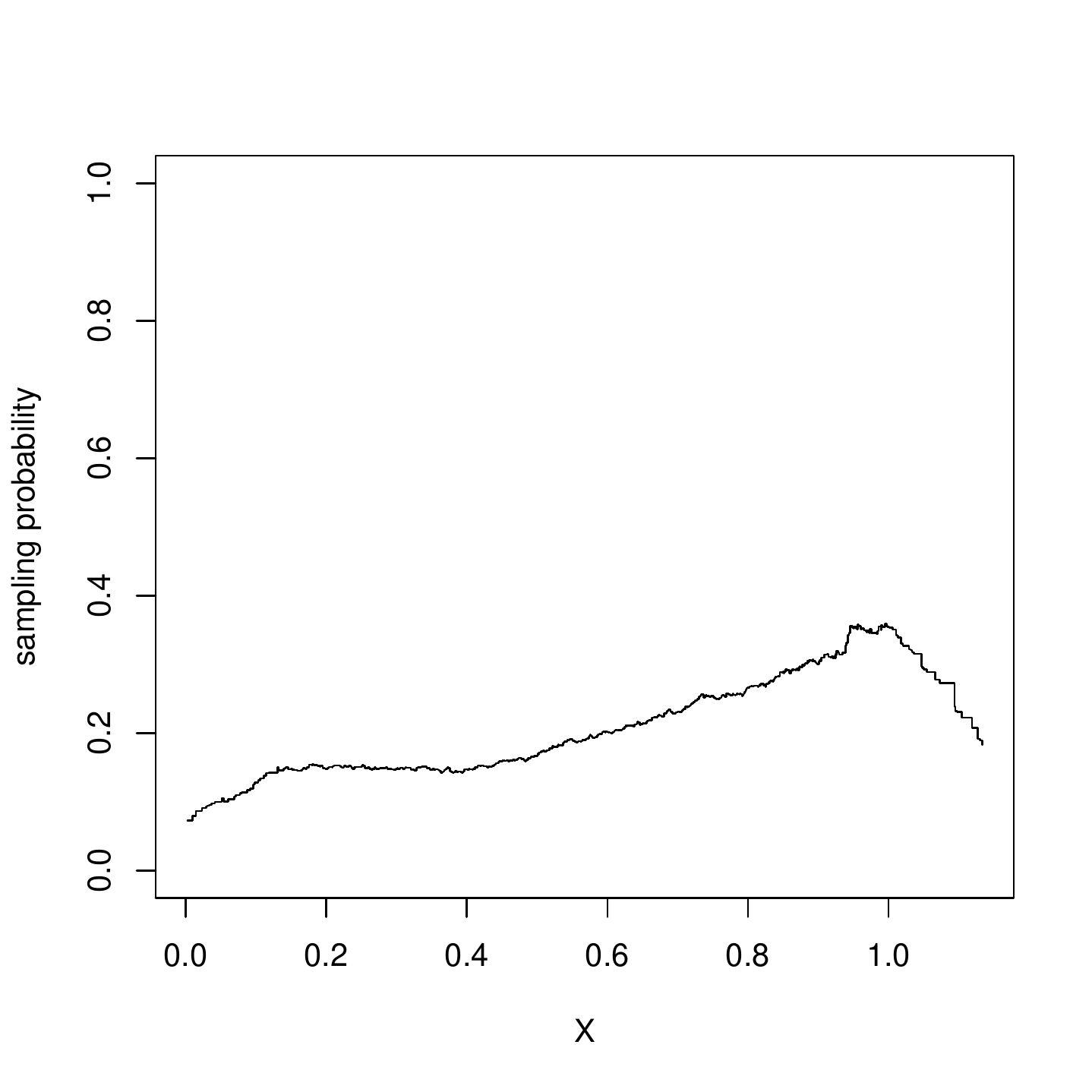}
	\caption{NPMLE of the sampling probability $G(x)$ based on a sample of size $n=1,000$ taken from the simulated Cox model.} %Due to the double truncation, large values of $X$ are observed with a relatively larger probability.
	\label{fig:G}
\end{figure}

\noindent Variables $(U,V)$ were generated independently of the target $(X,Z)$. The model (LT)-(RT) represents a situation with interval sampling and window width $0.25$. Note that the supports of $U$ and $V$ are the intervals $(-\tau,1)$ and $(0,1+\tau)$, respectively, so the choice $\tau=0.25$ is large enough for the identification of $F$, the cdf of $X$. On the other hand, under (LT)-(RT) the sampling probability $G(x)=P(U\leq x\leq V)$ is non constant, see Figure \ref{fig:G}; hence, the application of the naive estimator for $\beta$ which ignores the truncation will fail. The truncation rate in the simulations was about $81 \%$.\\

In Table \ref{tab:cox} the bias, SD and MSE of Mandel et al. (2018)'s and Rennert and Xie (2018)'s estimators along $250$ Monte Carlo trials are reported. For comparison purposes, the benchmark estimator without truncation as well as the naive approach which ignores the double truncation are considered too. From Table \ref{tab:cox} it is seen that both estimation approaches are virtually unbiased and consistent, and that the one in Mandel et al. (2018) is better since it reports estimations with smaller variances; the level of improvement of Mandel et al. (2018)'s over Rennert and Xie (2018)'s is larger for small sample sizes. It is also seen that the double truncation results in a loss of efficiency; for example, with $n=250$ the MSE relative to the benchmark method is $1.67$ and $2.10$ for \textit{man} and \textit{ren} approaches, these figures reducing to $1.47$ and $1.70$ with $n=1,000$. Finally, Table \ref{tab:cox} reveals the systematic bias of the naive estimator which does not correct for truncation.

\begin{table}[!ht]\centering
	\caption{\label{tab:cox} Bias, standard deviation (SD) and mean squared error (MSE) along $250$ trials of estimators for regression coefficients: benchmark estimator for no truncation (\textit{ben}), naive estimator which ignores the truncation (\textit{nai}), Mandel et al. (2018)'s method (\textit{man}), and Rennert and Xie (2018)'s method (\textit{ren}). Cox model. True regression coefficient is $\beta=1/\sigma=10$.}
	\medskip
	\begin{tabular}{cccccccccc}
		% First line:
	%	\toprule[0.09 em]
		% The body of the table:
		\hline
		 & & $n=250$ & & & $n=500$ & & & $n=1,000$ & \\
		 \hline
		 & Bias & SD & MSE   & Bias & SD & MSE  & Bias & SD & MSE \\
		\hline
	%	\midrule
		\textit{ben}  & .0273  & .5295 & .2811 & -.0083 & .4014 & .1612 & .0060 & .2412 & .0582 \\
		\textit{nai}  &  .4769 & .5885 & .5737 & .4731 & .4052 & .3880 & .4586 & .2451 &  .2704 \\ %
		\textit{man}  &  .0842 & .6802 & .4698 & .0601 & .4619 & .2169 & .0508 & .2879 & .0855 \\
		\textit{ren}  &  .0455 & .766 & .5893 & .0594 & .4928 & .2464 & .0588 & .3087 & .0988  \\
		% last line:
	%	\bottomrule[0.09 em]
	\hline
	\end{tabular}
\end{table}

\section{Illustrative data analyses}

In this Section applications with real doubly truncated data are provided in order to illustrate the capabilities and relative results of the reviewed packages. Estimation of the marginal cdf of the variable of interest $X$ by parametric and nonparametric methods, cumulative incidences for competing risks, Cox regression analyses and quasi-independence tests are included.

\subsection{Estimation of the marginal distribution}

In this section we focus on the estimation of the cdf for the age at onset of childhood cancer. The data refer $n=401$ children diagnosed with cancer in the region of North Portugal between January 1, 1999, and December 31, 2003; see de U\~{n}a-\'{A}lvarez (2020) for further details. As indicated in the Introduction, the age at onset $X$ is doubly truncated due to the interval sampling; specifically, if $V$ is the age (in days) by December 31, 2003, and $U=V-1825$, then the available data are restricted to $U\leq X\leq V$. The head of the dataset is as follows:

\begin{verbatim}
> my.child <- read.table("childcancer.cmprsk.txt", header = TRUE)
> head(my.child)
x     u    v z
1  6 -1643  182 4
2  7   -24 1801 2
3 15  -532 1293 4
4 43 -1508  317 4
5 85  -691 1134 6
6 92 -1235  590 5
\end{verbatim}

A test for quasi-independence between $X$ and $U$ was performed with the package \texttt{SurvTrunc} by running \texttt{indeptestDT(x, u, v)}. Kendall's Tau was $\tau_L=-0.01$ reporting a p-value of $0.952$, so the quasi-independence was accepted.\\

\begin{figure}[h]
	\centering
	\includegraphics[scale=0.7]{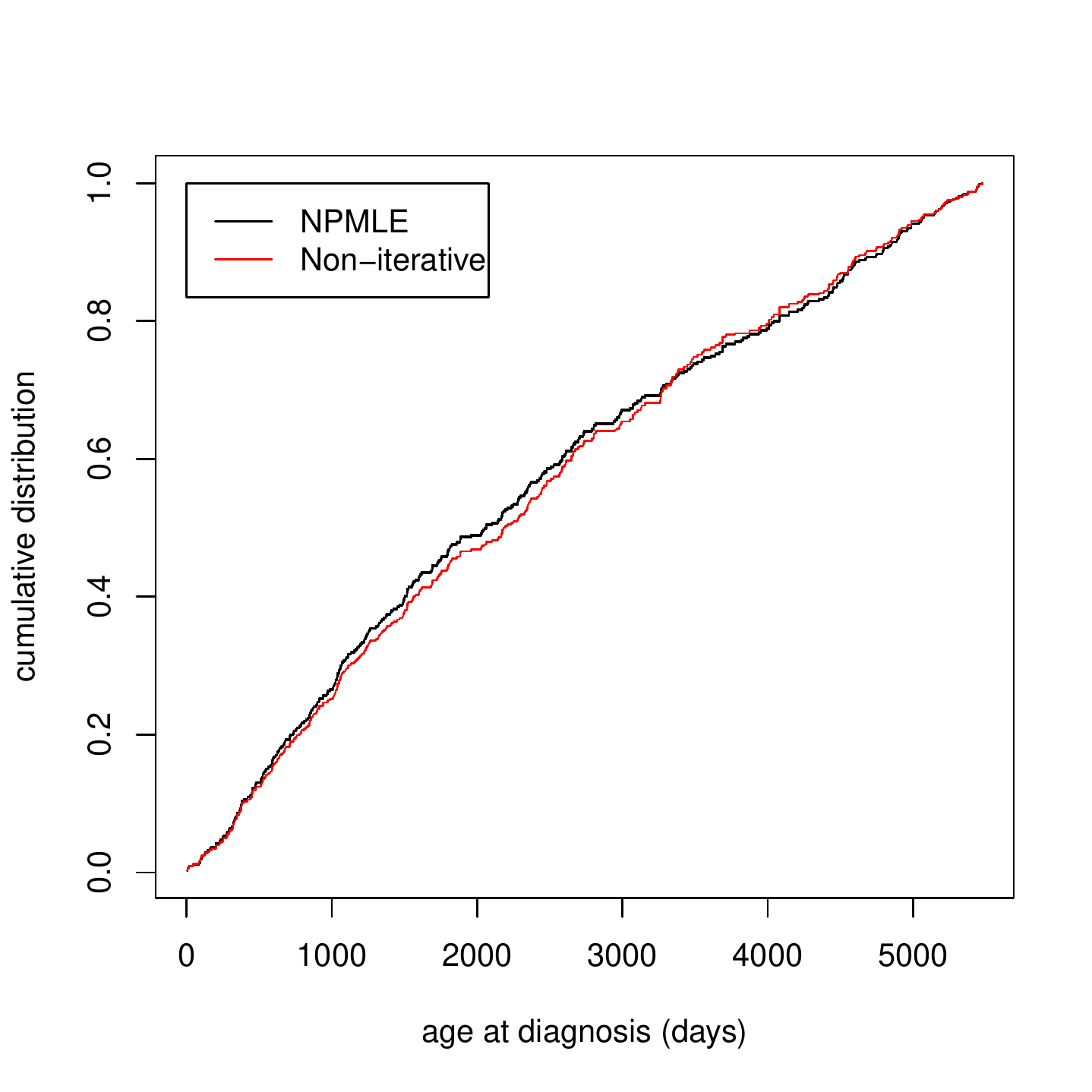}
	\caption{Nonparametric estimators of the age at diagnosis, childhood cancer data: NPMLE (black line) and non-iterative estimator (red-line).}
	\label{fig:marginal_nonp}
\end{figure}

Figure \ref{fig:marginal_nonp} displays the NPMLE of the cdf for the age at cancer diagnosis (black line), together with the non-iterative estimator (red line). The NPMLE was computed by using the six different \texttt{R} functions in the packages \texttt{DTDA}, \texttt{SurvTrunc}, \texttt{double.truncation} and \texttt{DTDA.cif} reviewed above; the results overlap. The running time was below 2 seconds except for \texttt{lynden} ($5.73$ seconds) and \texttt{shen} ($169.78$), revealing an inefficient implementation in the latter case. Code lines for Figure \ref{fig:marginal_nonp} were

\begin{verbatim}
> ep <- efron.petrosian(x, u, v, boot = FALSE)
> ly <- lynden(x, u, v, boot = FALSE)
> sh <- shen(x, u, v, boot = FALSE)
> st <- cdfDT(x, u, v)
> dt <- NPMLE(u, x, v)
> sh2 <- DTDAcif(x, u, v )
> tau <- (v - u)[1]
> ni <- DTDAni(x, u, tau)
> plot(ep$time, ep$cumulative.df, type = "s", ylab = "cumulative distribution",
     + xlab = "age at diagnosis (days)")
> lines(ly$time, ly$cumulative.df, type = "s", col = 1)
> lines(sh$time, sh$cumulative.df, type = "s", col = 1)
> lines(st$time, st$F, type = "s", col = 1)
> lines(x[order(x)], dt$F, type = "s", col = 1)
> lines(sh2$data$x, cumsum(sh2$cif.mas), type = "s", col = 1)
> lines(ni$x, ni$cumprob, type = "s", col = 2)
> legend(0, 1, legend = c("NPMLE", "Non-iterative"), col = c(1,2), lty = c(1,1))


\end{verbatim}

\begin{figure}[h]
	\centering
	\includegraphics[scale=0.7]{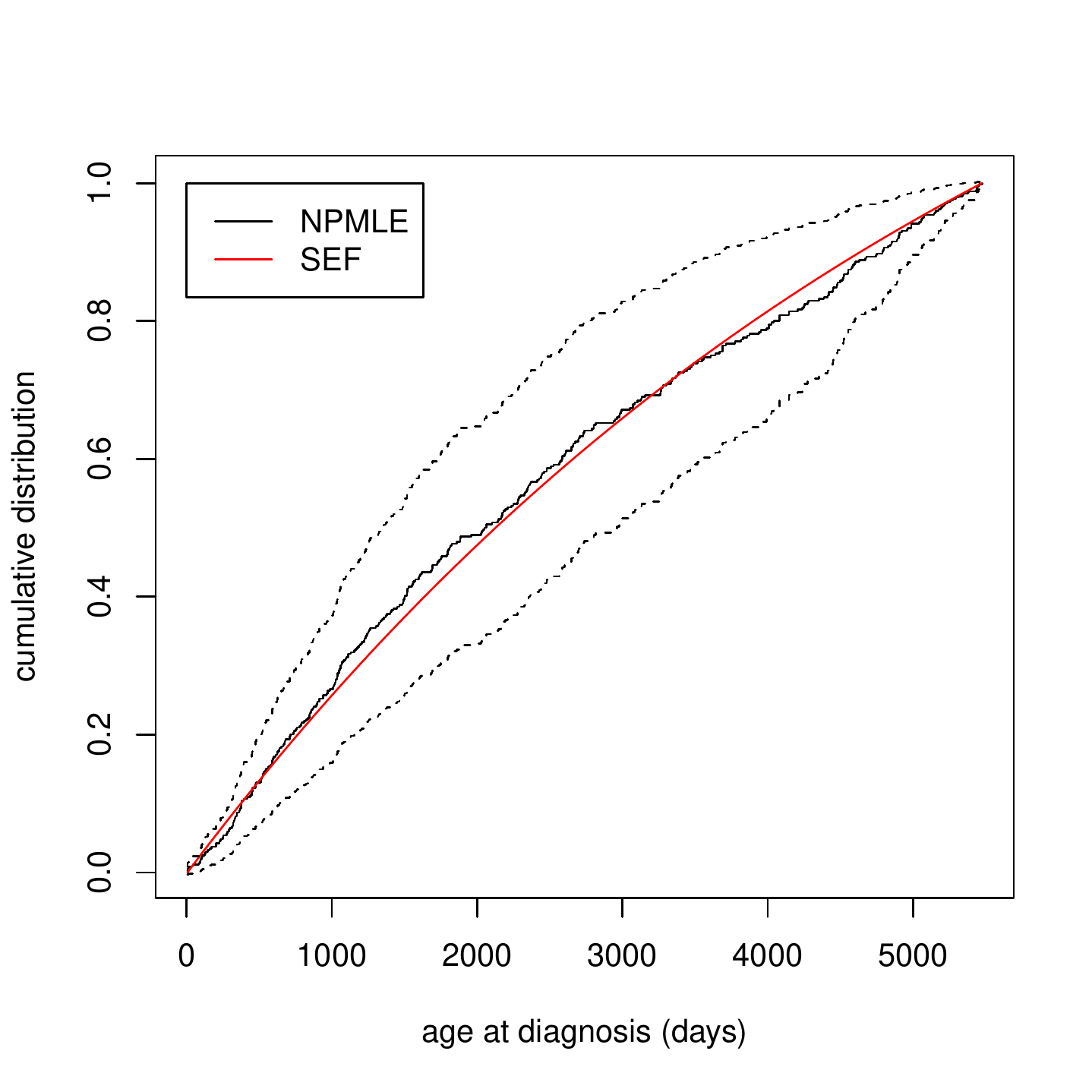}
	\caption{Estimators of the age at diagnosis, childhood cancer data: NPMLE with 95\% confidence limits (black lines) and parametric estimator based on the special exponential family (red-line).}
	\label{fig:marginal_SEF}
\end{figure}

The package \texttt{double.truncation} was used to select the special exponential family model with the smallest AIC, which was the one reported by the \texttt{PMLE.SEF1.free} function. The corresponding cdf was

\begin{equation}
F(x;\hat \eta)=1-\frac{\exp(\hat \eta \hat b_X)-\exp(\hat \eta x)}{\exp(\hat \eta \hat b_X)-\exp(\hat \eta \hat a_X)},\hspace{1.5 cm}\hat a_X<x<\hat b_X,
\label{eq:SEF}
\end{equation}

\noindent where $\hat a_X=\min(X_i)=6$, $\hat b_X=\max(X_i)=5474$, and $\hat \eta=-0.00017$. In Figure \ref{fig:marginal_SEF} the cdf (\ref{eq:SEF}) is displayed; for completeness, the NPMLE together with 95\% confidence limits (both computed through \texttt{NPMLE}) are included too. The following code was used to generate Figure \ref{fig:marginal_SEF}:

\begin{verbatim}
> par <- PMLE.SEF1.free(u, x, v)
> tau1 <- min(x)
> tau2 <- max(x)
> cdf <- 1- (exp(par$eta*tau2) - exp(par$eta*x)) /
     + (exp(par$eta*tau2) - exp(par$eta*tau1))
> plot(x[order(x)], dt$F, type = "s", ylab = "cumulative distribution", 
     + xlab = "age at diagnosis (days)")
> lines(x[order(x)], cdf[order(x)], type = "l", col = 2)
> upp <- dt$F - dt$SE*qnorm(.025)
> low <-dt$F + dt$SE*qnorm(.025)
> lines(x[order(x)], upp, type = "s", lty = 2)
> lines(x[order(x)], low, type = "s", lty = 2)
> legend(0, 1, legend = c("NPMLE", "SEF"), lty = 1, col = 1:2)
\end{verbatim}

%Introduce PD data here. Interesting because of the observational bias but, more importantly, because the strong rouding effects. It is seen that the result of \texttt{lynden} differs from the one reported by the other numerical implementations. My preliminary conclusion is that \texttt{lynden} does not properly compute the NPMLE with ties; also, the implementation of the non-iterative estimator in \texttt{DTDAni} could not be suitable in such a case.

\subsection{Estimation of cumulative incidences for competing risks}

\begin{figure}[h]
	\centering
	\includegraphics[scale=0.7]{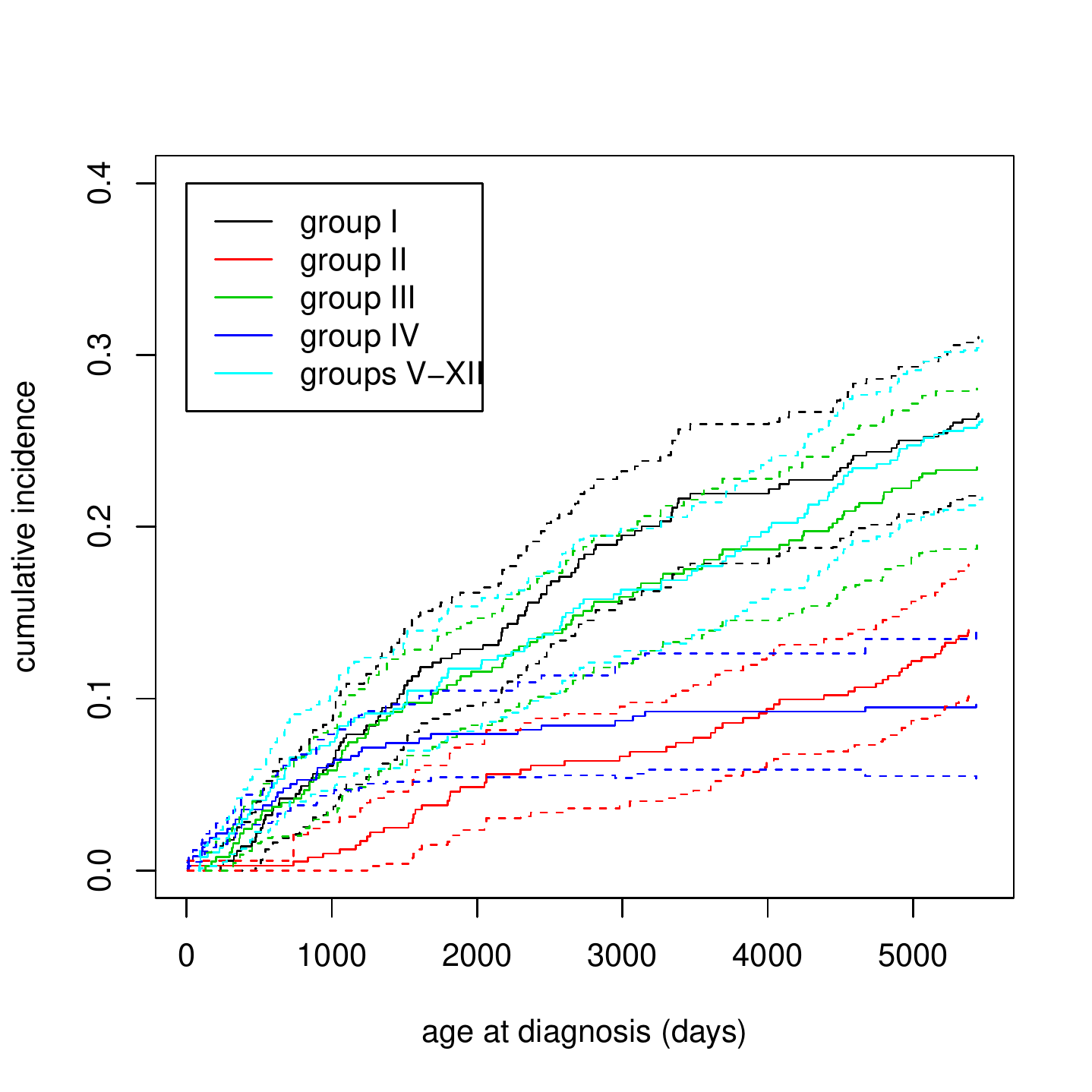}
	\caption{Nonparametric estimators of the cumulative incidences for the several ICCC groups, childhood cancer data. Dashed lines correspond to 95\% pointwise confidence limits based on $99$ bootstrap resamples.}
	\label{fig:cifs_indep}
\end{figure}

Several cancer
types are present in the childhood cancer registry. Cases were grouped according to the International Classification of Childhood Cancer (ICCC). Specifically,
the cancer groups are leukemias (group I, 107 cases), lymphomas (II, 57), central nervous system (III, 94), neuroblastoma (IV, 38), and other less
frequently observed cancers that were grouped together (ICCC groups V–-XII, 105 cases). In Figure \ref{fig:cifs_indep} the cumulative incidence for the several ICCC groups are displayed together with 95\% pointwise confidence limits based on $99$ bootstrap resamples. In order to get estimations with relatively small standard errors the method based on the independence assumption between the truncating variables and the cancer type was used. Here there are the code lines needed to get Figure \ref{fig:cifs_indep}:

\begin{verbatim}
> z[z>=5] <- 5
> cr.boot <- DTDAcif(x, u, v, z, method = "indep", boot = TRUE, B = 99)
> plot(cr.boot, ylab = "cumulative incidence", 
     + xlab = "age at diagnosis (days)", intervals = TRUE, ylim = c(0, 0.4))
> legend(0, .4, legend = c("group I", "group II",
     + "group III", "group IV", "groups V-XII"), lty = 1, col = 1:5)
\end{verbatim}

\subsection{Cox regression}

In this section Cox regression is applied to investigate the possible influence of two single nucleotide polymorphisms (SNPs) on the risk or age of onset of Parkinson's disease (PD). The setting is a study of the association of candidate SNPs and age of onset of PD (Clark et al., 2011). Following Clark et al. (2011), the rs8192678 PGC-1a SNP and the A10398G mitochondrial SNP are considered. As indicated by Mandel et al. (2018), in this study the selected patients are the ones with DNA sample taken within eight years of their onset of PD and, besides, the onset was required to be prior to the DNA sample. Therefore, the age of onset $X$ is right truncated by the age at blood sampling for genetic analysis $V$, and left truncated by $U=V-8$. We focus on the late onset group of patients ($n=100$), who had onset ages between $63$ and $87$ years; the genetic information of these cases is reported in Table \ref{tab:PD}.\\

\begin{table}[!ht]\centering
	\caption{\label{tab:PD} SNP distribution for the Parkinson's disease data -late onset group.}
	\medskip
	\begin{tabular}{ccccc}
		% First line:
		%	\toprule[0.09 em]
		% The body of the table:
		\hline
		  &  &  & PGC-1a &  \\
	      &   & A & AG & G  \\ %
		SNP10398  &  A & 7 & 30 & 36 \\
		 &  G & 3 & 7 & 17 \\
		% last line:
		%	\bottomrule[0.09 em]
		\hline
	\end{tabular}
\end{table}

The regression coefficients were estimated by two different methods: the one in Mandel et al. (2018), and the alternative approach proposed by Rennert and Xie (2018). For these two approaches standard errors based on the simple bootstrap with $199$ bootstrap resamples were obtained, and the corresponding Wald-type two-sided p-values were calculated. The results are displayed in Table \ref{tab:coxest}. The code lines were as follows:

\begin{verbatim}
> library(DTDA.cif)
> library(survival)
> library(SurvTrunc)
> late <- read.csv("pdlate_8_12_09.csv",header=T,sep=",")
> x <- late[,1]; u <- late[,2]-8; v <- late[,2]
> z1 <- late[,3]
> z2 <- late[,4]
> W <- DTDAcif(x, u, v)$biasf
> res.cox <- coxph(Surv(x) ~ z1 + z2 + offset(-log(W)))
> set.seed(1234)
> n <- length(x)
> B.man <- 199
> beta.man <- matrix(nrow = B.man, ncol = 3)
> for (b in 1:B.man){
+ 
+ i.man <- sample(n, replace = TRUE)
+ xb <- x[i.man]
+ ub <- u[i.man]
+ vb <- v[i.man]
+ z1b <- z1[i.man]
+ z2b <- z2[i.man]
+ Wb <- DTDAcif(xb, ub, vb)$biasf
+ res.coxb <- coxph(Surv(xb) ~ z1b + z2b + offset(-log(Wb)))
+ beta.man[b, ]  <- coef(res.coxb)
+ 
+ }
> se.man <- apply(beta.man, 2, sd)
> p.man <- 2 * pnorm(-abs(coef(res.cox)/se.man))
> res.cox2.199 <- coxDT(Surv(Onset.Age) ~ X10398 + 				
     PGC1A_GLY482SER, Sampling.Age - 8, Sampling.Age,
     data = late, B.SE.np = 199)
\end{verbatim}

\begin{table}[!ht]\centering
	\caption{\label{tab:coxest} Estimated regression coefficients, bootstrap standard errors and Wald-type two-sided p-values for the Parkinson's disease data -late onset group. Results correspond to Mandel et al. (2018)'s approach and Rennert and Xie (2018)'s method -figures for the latter in brackets.}
	\medskip
	\begin{tabular}{ccccc}
		% First line:
		%	\toprule[0.09 em]
		% The body of the table:
		\hline
		SNP & Estimate & Std.err  & p-value \\
		\hline
		SNP10398 G & 0.5980 (-.1776) & .3061 (.3200) & .0508 (.5789)   \\ %
		PGC-1a AG  & -1.1926 (.4306)  & .8793 (.5684) & .1750 (.4487) \\
		PGC-1a G & -.6383 (-.0419) & .3226 (.3673) & .0478 (.9091) \\
		% last line:
		%	\bottomrule[0.09 em]
		\hline
	\end{tabular}
\end{table}

From Table \ref{tab:coxest} it is seen that the results provided by the two estimation approaches are quite different. For example, with Mandel et al. (2018)'s approach, at 5\% level SNP10398 almost reaches significance, while one concludes that PGC-1a changes the risk of PD. This is in well agreement with the results in Mandel et al. (2018), Table 4, in which confidence intervals based on the percentile bootstrap are reported. In contrast to this, the application of the method proposed by Rennert and Xie (2018) does not allow to get any significance. This seems to be due not only to the different standard errors, which tend to be larger for Rennert and Xie (2018)'s approach, but also to a shift in the point estimates.\\

\begin{figure}[htp]
	\centering	
	
	%\caption{equation...}	
	\begin{tabular}{cc}		
		\includegraphics[width=60mm]{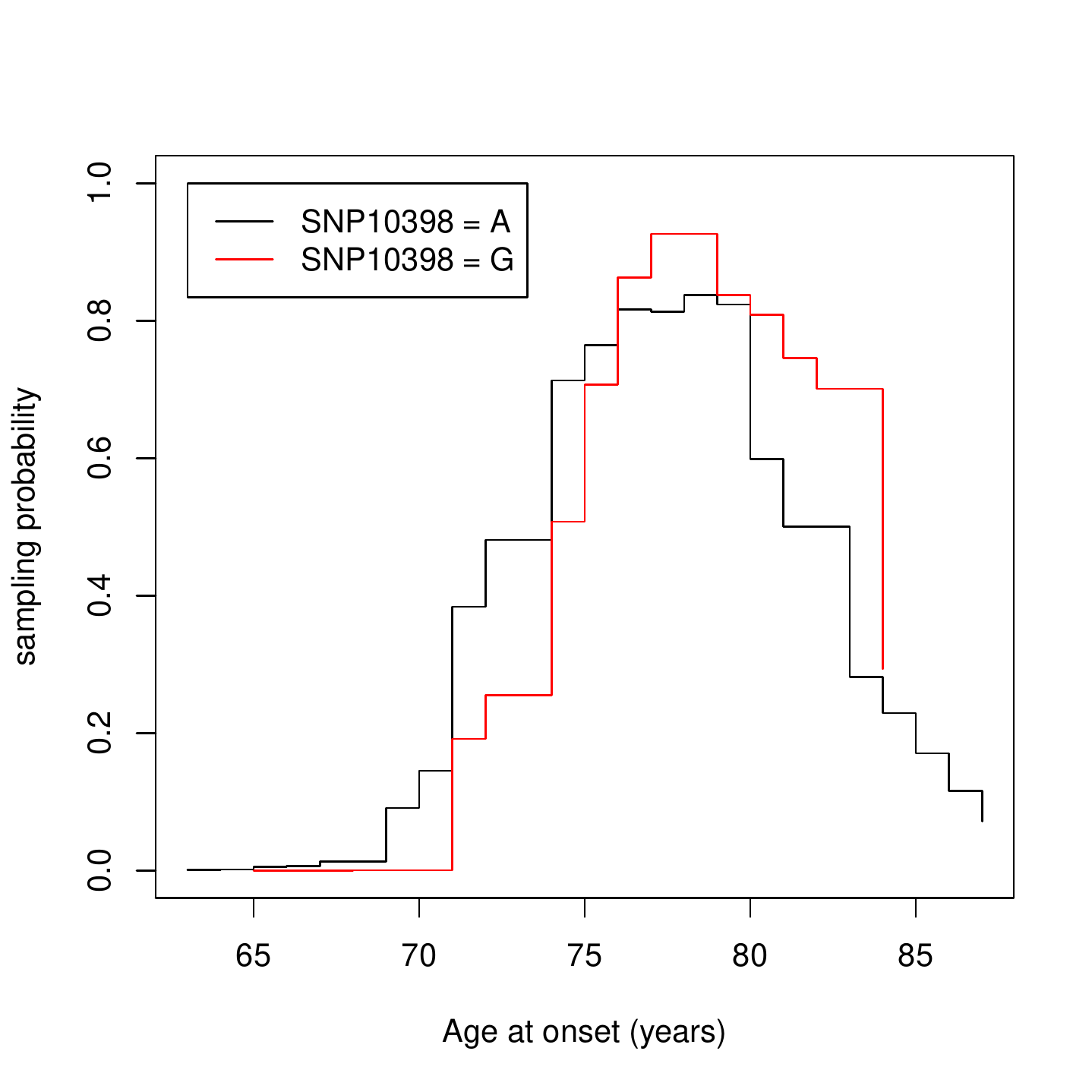}&	
		\includegraphics[width=60mm]{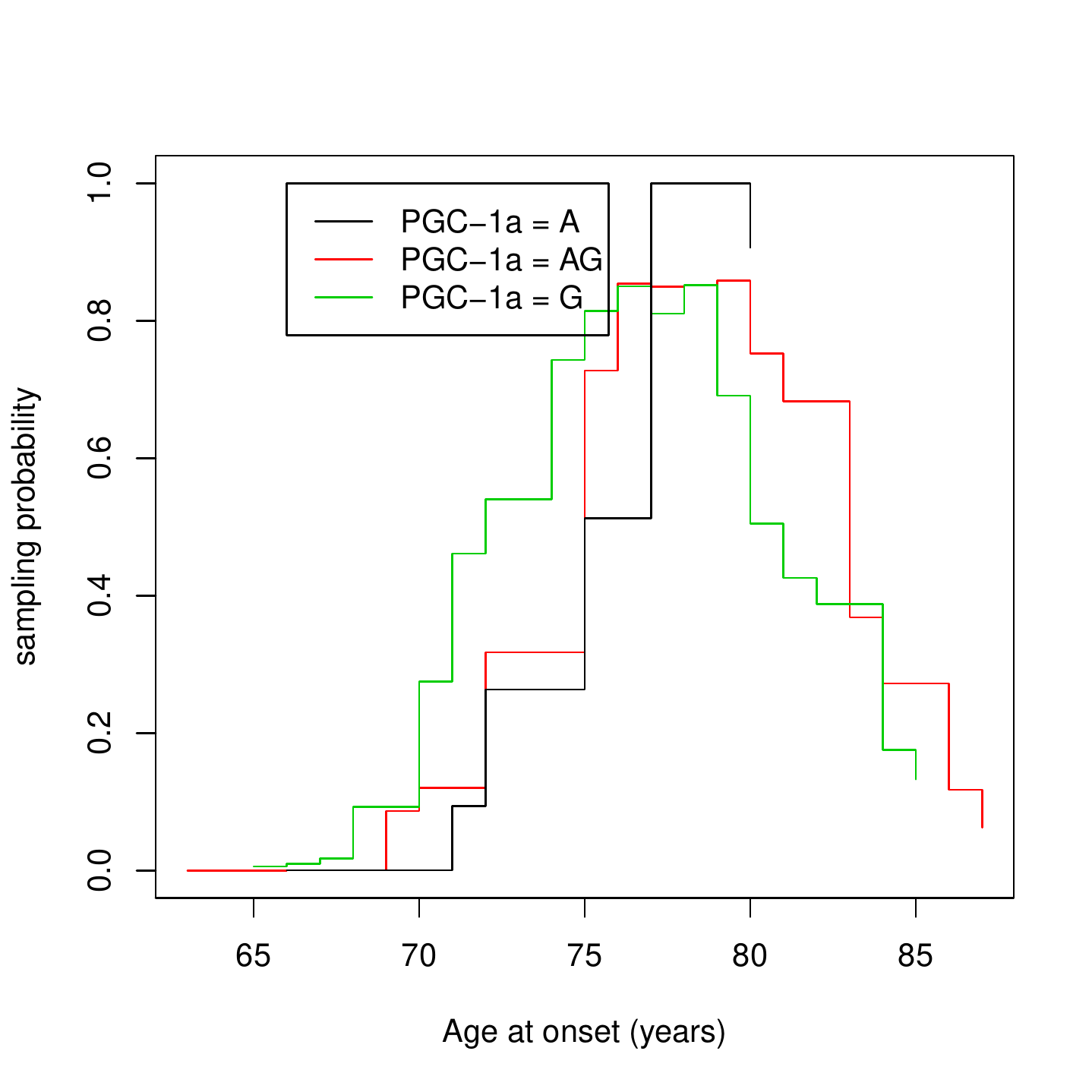}\\		
	\end{tabular}
	\caption{Sampling probabilities for the age at onset of PD depending on the genetic information -late onset group.}	
	\label{fig: samplingprobPD}
\end{figure}

The two estimation approaches in Table \ref{tab:coxest} require the quasi-independence between the age at onset of PD and the truncating interval, as well as the assumption that the sampling probability $G(x|z)=P(U \leq x\leq V|Z=z)$ is free of the covariate vector $Z$. The application of the \texttt{indeptestDT\{SurvTrunc\}} reported a Kendall's Tau of $\tau_L=0.158$ with p-value $0.103$, thus indicating no violation of the quasi-independence condition. On the other hand, in Figure \ref{fig: samplingprobPD} the sampling probabilities for the age at onset of PD along the several groups of individuals are displayed. The curves are close to each other, suggesting that $G(x|z)$ is not strongly influenced by the particular $z$-value.

\section{Discussion}

The statistical analysis of doubly truncated data is non-trivial. Luckily, several \texttt{R} libraries implementing methods for handling the double truncation exist. This facilitates the application of suitable corrections of standard procedures which can remove the potential sampling bias. In particular applications one will come up with a roughly flat curve $G_n(x)$, suggesting that the impact of double truncation is minor. This occurs for example in the childhood cancer data study of Section 5. In such circumnstances the user may decide to apply ordinary methods, which will report valid point estimates and standard errors under the null assumption '$G(x)$ is constant'. Alternative, he/she may keep the estimator $F_n(x)$ which corrects for double truncation together with its standard error to recognize the initial uncertainty on $G(x)$. In the latter case the variance will be larger and significant features will be more hardly found.\\

The reviewed packages implement procedures which require the independence (or quasi-independence) between the targeted variable and the truncation couple. A test for the quasi-independence assumption is implemented in \texttt{SurvTrunc} and should be applied as a first step. In case of rejection, alternative procedures which take the dependence structure between $X$ and $(U,V)$ into account are available; see e.g. Moreira et al. (2018). Also, in the regression setting, the function $G(x|z)$ should be free of the specific covariate value $Z=z$. This can be explored by graphically displaying the estimator $G_n$ for several $Z$-groups. No formal test for the independence between $(X,Z)$ and $(U,V)$ has been implemented so far.\\

The method implemented by \texttt{DTDAcif} computes the solution to the score equations in Shen (2010), thus allowing for the estimation of $F$ and of the observational bias $G$. This follows the spirit of the functions \texttt{shen\{DTDA\}} and \texttt{cdfDT\{SurvTrunc\}}. However, \texttt{DTDAcif} improves the computational speed by reducing the number of targets to a minimum and discarding the construction of automatic plots.  Specifically, the simplified algorithm in \texttt{DTDAcif} iterates between the current solutions for $G_n$ and $F_n$, by using the couple of equations

\begin{itemize}
	\item [(i)] $F_n(x)=\sum_{i=1}^n w_i(G_n)I(X_i \leq x) / \sum_{i=1}^n w_i(G_n)$,
	\item [(ii)] $G_n(x)=\sum_{i=1}^n w_i(F_n) I(U_i\leq x\leq V_i) / \sum_{i=1}^n w_i(F_n)$
	
\end{itemize}

\noindent where $w_i(G_n)$ and $w_i(F_n)$ are as in (S1)-(S2), section 2. Unlike for \texttt{shen} and \texttt{cdfDT}, the estimator for the truncation distribution is not returned. More importantly \texttt{DTDAcif} speeds the process up by implementing fast execution modules based on \texttt{C++}. Summarizing, \texttt{DTDA.cif} is the fastest package when both $F$ and the sampling probabilities are of interest. The execution times of \texttt{DTDAcif} may vary when the analysis includes competing events (\texttt{comp.event} argument) and \texttt{method = "dep"} is selected, since in such a case computations are performed in a slightly different way, more related to the implementations in \texttt{shen\{DTDA\}}.\\

When the sampling probabilities $G_n(X_i)$ are not required, the function \texttt{efron.petrosian\{DTDA\}} is recommended because of its computational speed; this function is only beated by \texttt{DTDAni}, which provides a non-iterative approximation to the NPMLE. However, if standard errors are to be computed, \texttt{NPMLE\{double.truncation\}} is quite faster than the functions which implement the bootstrap (like \texttt{efron.petrosian}). Still, the statistical accuracy of the explicit-form standard error in \texttt{NPMLE} can be poor relative to that of the bootstrap, particularly for small sample sizes ($n \leq 100$). Some evidence on this has been provided in Section 4, but more investigation is needed in order to reach general conclusions. For the moment some caution on this regard is indicated. Some errors leading to the non-availability of the standard error were found when running \texttt{NPMLE} too, being particularly frequent in the case $n=50$. So this is an extra drawback of the implementation of the explicit-form standard error which should be taken into account in applications. Regarding Cox regression, the approach in Mandel et al. (2018) is recommended over the one implemented in \texttt{coxDT\{SurvTrunc\}} due to its better statistical precision. The former can be easily performed by including the transformed sampling probabilities as an \texttt{offset} in the function \texttt{coxph\{survival\}}. The reported standard errors must be discarded and recalculated by bootstrapping.\\

Importantly, we recall that specific techniques may be available in a unique package. This occurs for instance with the cumulative incidences for competing risks (\texttt{DTDA.cif}) or with the estimation of a parametric model for the cdf of $X$ (\texttt{double.truncation}). Also, we mention that there exist relevant procedures for doubly truncated data which, at the present date, have not been implemented in any software package. This is the case, for example, of the smoothers for the density and regression functions introduced in Moreira and de U\~{n}a-\'{A}lvarez (2012) and Moreira et al. (2016) respectively; or of the estimators for the regression coefficients in the accelerated failure time model investigated by de U\~{n}a-\'{A}lvarez and Van Keilegom (2019). In principle, however, such methods can be applied in an easy way by including the inverse sampling probabilities $G_n(X_i)^{-1}$ in the \texttt{weights} option of the \texttt{R} functions \texttt{density}, \texttt{loess} or \texttt{lm}; when doing so, reported standard errors should be corrected by \textit{e.g.} bootstrapping to take the random weights into account. The application of procedures which involve the computation of new weights, for instance the one in Moreira et al. (2018) for dependent truncation, is less trivial.\\

Finally, problems with the possible non-existence or non-uniqueness of the NPMLE were recently pointed out by Xiao and Hudgens (2019). For example, a necessary condition for the existence and uniqueness of $F_n$ is that both $S_{1i}\equiv \sum_{k=1}^n I(U_k\leq X_i\leq V_k)$ and $S_{2i}\equiv \sum_{k=1}^n I(U_i\leq X_k\leq V_i)$ are strictly greater than 1 for $1\leq i \leq n$. We have verified that this condition holds for the two real datasets in Section 5. Interestingly, for the childhood cancer registry, it happened $\min_i S_{2i}=1$ when restricting the analysis to ICCC group III (central nervous system); indeed, \texttt{DTDAcif} fails to report valid estimates when setting \texttt{method = "dep"}, because with this option the NPMLE for each cancer group is computed.  Other, more sophisticated data checks can be performed following the results in Xiao and Hudgens (2019). To sum up, the issue of non-existence or multiplicity of the NPMLE deserves serious attention
in practice.\\

% but the standard errors are needed, the function \texttt{NPMLE\{double.truncation\}} is recommended because of its computational speed.

% Besides, even when this function does not provided an estimation of $G_n$, this can be easily obtained from the returned $F_n$ by using equation (ii) above. In order to illustrate the computational savings, the efficiency of \texttt{efron.petrosian} with 99 bootstrap resamples relative to \texttt{NPMLE} (measured as the ratio of the corresponding execution times) is depicted in Figure \ref{fig:RCE}.\\

%Text here.

%\begin{figure}[h]
%	\centering
%	\includegraphics[scale=0.7]{rel_comput_effic.pdf}
%	\caption{Computational efficiency of \texttt{efron.petrosian} (with 99 bootstrap resamples) relative to \texttt{NPMLE} for a single trial of the model in Table \ref{tab:exec_noboot}, case $\rho=0.5$.}
%	\label{fig:RCE}
%\end{figure}

%\begin{figure}[htp]
%	\centering	
	
	%\caption{equation...}	
%	\begin{tabular}{cc}		
		%\includegraphics[width=60mm]{comput_effic.pdf}&	
		%\includegraphics[width=60mm]{rel_comput_effic.pdf}\\		
	%\end{tabular}
	%\caption{Left: Execution time (minutes) of \texttt{efron.petrosian} with 99 bootstrap resamples and \texttt{NPMLE} depending on the sample size. Right: Computational efficiency of \texttt{efron.petrosian} with 99 bootstrap resamples relative to \texttt{NPMLE}. Single trial of the model in Table \ref{tab:exec_noboot}, case $\rho=0.5$. }	
	%\label{fig: CE}
%\end{figure}

\section*{Computational details}

The results in this paper were obtained using
\texttt{R}~3.6.3. \texttt{R} itself and all packages used are available from the
Comprehensive \texttt{R} Archive Network (CRAN) at https://CRAN.R-project.org/.

\section*{Acknowledgments}

Work supported by the Grant MTM2017-89422-P (MINECO/AEI/FEDER, UE). Financial support from the Xunta de Galicia (Centro singular de investigaciónn de Galicia accreditation 2019-2022) and the EU (ERDF), Ref. ED431G2019/06, is acknowledged too. Fruitful discussions with Carla Moreira, Rosa Crujeiras and Jos\'{e} Carlos Soage are acknowledged.

\section{References}

Clark J, Reddy S, Zheng K, Betensky RA, and Simon
DK (2011) Association of PGC-1alphapolymorphisms
with age of onset and risk of Parkinson’s disease. BMC Medical Genetics \textbf{12}, 69.\\

de U\~{n}a-\'{A}lvarez J (2018) A non-iterative estimator for interval sampling and
doubly truncated data. In: The Mathematics of the Uncertain –A Tribute to
Pedro Gil (Gil E, Gil E, Gil J, Gil MA Eds.) Studies in Systems, Decision and
Control, vol. 142. Springer, Cham, pp. 387--400.\\

de U\~{n}a-\'{A}lvarez J (2020) Nonparametric estimation of the cumulative incidence of competing risks under double truncation. Biometrical Journal 2020, 1--16. https://doi.org/10.1002/bimj.201800323\\

de U\~{n}a-\'{A}lvarez J, Soage JC (2018) DTDA.ni: Doubly Truncated Data Analysis, Non Iterative.
R package version 1.0.\\

de U\~{n}a-\'{A}lvarez J, Soage JC (2020) DTDA.cif: Doubly Truncated Data Analysis, Cumulative
Incidence Functions. R package version 1.0.2.\\

de U\~{n}a-\'{A}lvarez J, Van Keilegom I (2019) Efron-Petrosian integrals for doubly truncated data with covariates: an asymptotic analysis. Under review.\\

D\"{o}rre A (2017) Bayesian estimation of a lifetime distribution under double truncation caused by time-restricted data collection. Statistical Papers 2017. https://doi.org/10.1007/s00362-017-0968-7\\

Efron B, Petrosian V (1999) Nonparametric methods for doubly truncated data. Journal of the American Statistical Association 94, 824--834.\\

Emura T, Konno Y, Michimae H (2015) Statistical inference based on the nonparametric maximum
likelihood estimator under double-truncation. Lifetime Data Analysis 21, 397--418.\\

Hu YH, Emura T (2015) Maximum likelihood estimation for a special exponential family under
random double-truncation, Computational Statistics 30, 1199--1229.\\

Emura T, Hu YH, Huang CY (2019) double.truncation: Analysis of Doubly-Truncated
Data. R package version 1.4.\\

Emura T, Hu YH, Konno Y (2017) Asymptotic inference for maximum likelihood estimators under
the special exponential family with double-truncation. Statistical Papers 58, 877--909.\\

Mandel M, de U\~{n}a-\'{A}lvarez J, Simon DK, Betensky RA (2018) Inverse probability weighted cox regression for doubly truncated data.
Biometrics 74, 481–-487.\\

Martin EC, Betensky RA (2005) Testing quasi-independence of fauilure and truncation times via conditional Kendall's Tau. Journal of the American Statistical Association 100, 484--492.\\

Moreira C, de U\~{n}a-\'{A}lvarez J (2010) Bootstrapping the NPMLE for doubly truncated data. Annals of the Institute of Statistical Mathematics 22, 567--583.\\

Moreira C, de U\~{n}a-\'{A}lvarez J (2012) Kernel density estimation with doubly truncated data. Electronic Journal of Statistic 6, 501--521.\\

Moreira C, de U\~{n}a-\'{A}lvarez J, Braekers R (2018) Nonparametric estimation of a distribution function from doubly truncated data under dependence.
arXiv:1802.08579.\\

Moreira C, de U\~{n}a-\'{A}lvarez J, Crujeiras R (2010) \texttt{DTDA}: an \texttt{R} package to analyze randomly truncated data. Journal of Statistical Software 37(7), 1--20.\\

Moreira C, de U\~{n}a-\'{A}lvarez J, Crujeiras R (2020) DTDA: Doubly Truncated Data Analysis.
R package version 2.1.2.\\

Moreira C, de U\~{n}a-\'{A}lvarez J, Meira-Machado L (2016) Nonparametric regression with doubly truncated data. Computational Statistics and Data Analysis 93, 294--307.\\

R Core Team (2017) R: A Language and Environment for Statistical Computing. R Foundation
for Statistical Computing, Vienna, Austria.\\

Rennert L (2018) SurvTrunc: Analysis of Doubly Truncated Data. R package version 0.1.0.\\

Rennert L, Xie SX (2018) Cox regression model with doubly truncated data. Biometrics 74, 725--733.\\

Shen, PS (2010) Nonparametric analysis of doubly truncated data. Annals of the Institute of Statistical Mathematics 62, 835--853.\\

Xiao J, Hudgens MG (2019) On nonparametric maximum likelihood estimation with double truncation. Biometrika 106, 989--996.\\

Ye ZS, Tang LC (2016) Augmenting the unreturned for field data with
information on returned failures only. Technometrics 58, 513--523.\\

Zhu H, Wang MC (2014) Nonparametric inference on bivariate survival data with interval sampling: association estimation and testing. Biometrika 101, 519--533.

\end{document}